\documentclass[12pt]{article}
\newtheorem{theorem}{Theorem}
 
\newtheorem{definition}{Definition}

\usepackage{graphicx,color}
\usepackage{amsmath,amssymb,amsfonts,bm,slashed,graphics}

\usepackage{physics}
\usepackage{caption}
\usepackage{subcaption}

\newcommand{\bea}{\begin{eqnarray}}
\newcommand{\eea}{\end{eqnarray}}

\def\gep{{\epsilon}}\newcommand{\MM}{{\mathfrak M}}

\def\N{{\mathbb N}}

\def\R{{\mathbb R}}
\newcommand{\TT}{{\mathfrak T}}
\newcommand{\CD}{{\mathbf{\mathcal D}}}
\newcommand{\D}{{\mathcal D}}
\newcommand{\eps}{\epsilon}

\newcommand{\CL}{{\cal{L}}}

\newcommand{\CS}{{\mathcal S}}

\newcommand{\CB}{{\mathbf{\mathcal B}}}
\newcommand{\K}{{\mathfrak K}}
\newcommand{\CR}{{\mathcal R}}

\begin{document}

\title{The Tensor Track VIII: Stochastic Analysis}
\author{V. Rivasseau\\
Universit\'e Paris-Saclay, CNRS/IN2P3\\ IJCLab, 91405 Orsay, France}
\date{} 
\maketitle

\begin{abstract}Assuming some familiarity with quantum field theory and with the tensor track approach 
that we presented in the previous series “Tensor Track I-VII”, we provide, as usual, the developments in tensors models
of the last two years. Then we expose the fundamental breakthrough of Martin Hairer on regularity structures and the work of 
Léonard Ferdinand on stochastic analysis applied to super-renormalizable tensor field theories.
We conclude with the hope that this work could be extended to just-renormalizable and asymptotically free models.

\end{abstract}

\noindent\textbf{keywords} Quantum Gravity, Tensors Models, Stochastic Analysis.

\section{Introduction}

Random tensors, like matrix models, originated in theoretical physics.  
In the 70's the hot stuff in theoretical physics was to quantize the elementary particles like quarks and gluons.
In this period matrix models had a lot of success in quantizing the strong interaction.

\medskip
In the 90's the dominating theory in quantizing gravity was string theory.
Random  matrix  models were seen at this time as a successful theory for quantizing gravity, but only in two dimensions. 
The inventors of random tensor models, such as Ambjorn, Gross and Sasakura, wanted to replicate the success of matrix models for dimensions three and four. But they lacked an essential tool, the $1/N$ expansion.

\medskip
Let's come to 2010's. The tensor track \cite{tensortrack1}-\cite{tensortrack7} is an attempt to quantize gravity in dimensions greater than two, by combining random tensor models,  discrete geometry and the  renormalisation  group. The tensor track lies at the crossroad of  several  closely  related  approaches to quantize gravity,  
most  notably causal dynamical  triangulations,  
quantum field theory on non-commutative spaces and  group field theory. 
Random tensors share with random matrices the fact that they are a zero-dimensional world, and, as such, they are background-independent; they made no references whatsoever of any particular space-time. 

\medskip
Moreover, random tensors models, based on the quantum field theory of Feynman, are manageable by renormalisation group techniques. Simple just-renomalizable  models even share with non-Abelian gauge theories the property of asymptotic freedom. The simplest such model is the  $T^4_5$ model.

\medskip
Random tensors are expected to play a growing role in many areas of mathematics, physics, and computer science,
but communities using random tensors have grown apart, developing their own tools and results;
for an up-to-date review of distinct approaches to quantum gravity 
exposing shared challenges and common directions, we suggest consulting \cite{Boer}.

\section{The Tensor Track}
In \cite{tensortrack1}-\cite{tensortrack4} we proposed a new way of looking at the problem of quantum gravity. 
Let us summarize briefly what it's all about. 
First we would like to say that the tensor  track has its birth in extending the matrix models 
and their $1/N$ expansion to tensor models.

\medskip
In the Hermitian matrix ensemble (GUE) perturbed by a quadratic interaction,  the $1/N$ expansion is well known. 
The free partition function is $ \int d M e^{- \frac{N}{2}\Tr M^2 } $, where 
\begin{equation}
\quad d M= \prod_k d M_{kk} \prod_{i<j} d \Re M_{ij}d \Im M_{ij},
\end{equation}
and the expectation values of $U(N)$ invariants 
\begin{equation}
< \Tr M^{p_1} \Tr M^{p_2}...\Tr M^{p_k} >
\end{equation}
 is entirely determined by the propagator 
\begin{equation} C_{ij,kl} =  \frac{1}{N} \delta_{il}\delta_{jk}
\end{equation}
and by Wick's rule.
 
Any scalar function of a tensorial quantum field theory  
can be further decomposed as a big functional integral
on a Gaussian measure and an interactive part. 
In the tensorial case this interactive part is a sum of invariants of the tensor.
For example the partition function is a scalar function of $N$, defined by
\begin{equation}
Z(N) =  \int d \mu (T) e^{- \sum_{Inv} S_{Inv}( T) }.
\end{equation}
The partition function and the corresponding free energy (also a scalar function) are related by a normalized  logarithm 
\begin{equation}
F(N) =\frac{1}{N^D} \log Z(N) .
\end{equation}

\begin{figure}[!htb]
\centering
\includegraphics[width=12cm]{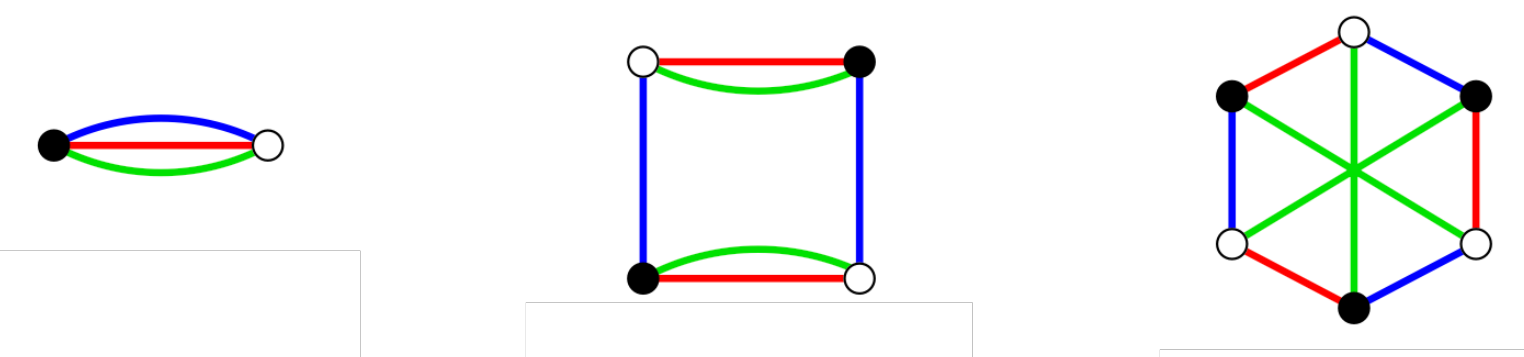} \caption{Examples of $U(N)$ invariants. The graphs presented in the left and center panels are \emph{melonic}, while the one on the right panel is not.}
  \label{coninv1}
\end{figure}

The invariants themselves can be classified in terms of graphs. Of course these graphs depend upon the group symmetries of the tensor. For matrix models 
the expectation values of the invariants can be classified by ribbon diagrams.
In the case of tensor models  Figure \ref{coninv1} depicts a partial list of connected invariants for $\bigotimes_{i=1}^3 U(N)$.
To generalize the $1/N$ expansion to the tensor case, the first step is to choose an invariant, for example a quartic invariant, and to normalize the $S_{Inv_0}(T)$ for that invariant:
\begin{equation}
S_{Inv_0}( T) \to  \frac{\lambda}{N^\alpha} S_{Inv_0}( T) .
\end{equation}
Now the partition function and the corresponding free energy depend on {\it  two variables}. In a quantum field theory the usual form of perturbation theory is to expand in power series in $\lambda$, and the perturbation is indexed by Feynman amplitudes associated to Feynman graphs.
For instance the perturbation of the free energy is of the following form  
\begin{equation}
F(\lambda,N)= \sum_{G} \frac{(-\lambda)^{v(G)} }{sym(G)}A ( G,N) .
\end{equation}
Once this is done, the hard step is to find $\alpha$ such that 1/N expansion exists, i.e.
such that the perturbation of the free energy is of the following form  
\begin{equation}
F(\lambda,N)= \sum_{G} \frac{(-\lambda)^{v(G)} }{sym(G)}A ( G,N) = \sum_{\omega  \in {\mathbb N} } 
N^{- \omega} F_{\omega} (\lambda ) .
\end{equation}

In the case of  $\bigotimes_{i=1}^D U(N)$ and a quartic $D$-melonic invariant,  
the $1/N$ expansion is governed by the Gur\u{a}u degree  and 
$F_{0} (\lambda ) $ is formed by all the $D+1$ melonic graphs (the 0 color being  associated to the Feynman propagators \cite{BonGurRiv}).
The family of $D+1$ melonic graphs \cite{BonGurRieRiv} which lead  the 1/N expansion 
of random tensors models can perhaps be called too trivial from a topologist point of view;
 it corresponds to some triangulations of the sphere $S_D$.
But, as the Gur\u{a}u degree is not only purely topological, the interplay between combinatorics and topology in   
sub-leading terms can be amazingly complex.

Now that we have been able to identify the leading terms in the $1/N$ expansion, 
the second step is to be able to resum them, i.e. to explicitly 
compute $F_{0} (\lambda ) $. This step has been performed for the first time by a paper by Gur\u{a}u and Ryan \cite{GurRya}.
From a probabilistic and statistical mechanical point of view, it corresponds to Aldous phase of branched polymers. 

Once we have been able to compute explicitly $F_{0} (\lambda) $, many possibilities are open to us:
 modifying the symmetry of the main tensor,  include the renormalization group by modifying 
in a specific way our propagators, a non-perturbative treatment of some simple models...

From the perspective of matrix models, to go further in the two parameters approach 
require a particular technique, namely double scaling. 
The first step of applying this technique to tensors has been done in \cite{DarGurRiv,GurSch}.
 The initial papers have been followed by mixed results, some results suggest the universality of branched polymers, 
 others pointing to the fact that some simple and natural restrictions change that universality class.
But, from the perspective of quantum gravity, the main goal is to resum the sub-leading terms
in order to find a more interesting phase of geometry pondered by Einstein-Hilbert action.

Let us come to the Sachdev-Ye-Kitaev (SYK) model. Discovered by Kitaev \cite{Kit}, it is a quartic model of $N$ Majorana fermions coupled by a disordered tensor  \cite{MalSta}. It is a model of condensed matter, 
hence it depends on time though a Hamiltonian. The disordered tensor is centered Gaussian iid
\begin{equation}
< J_{abcd} > = 0 ,  \quad < J^2_{abcd} >= \frac{\lambda^2}{N^3},
\end{equation}
and the Hamiltonian  is simply $H = J_{abcd} \psi_a \psi_b \psi_c \psi_d$.
This model posses three important properties: it is solvable at large $N$, there is a conformal symmetry at strong coupling, hence it can be a fixed point of the renormalization group,
 and, above all from quantum gravity, it is maximally chaotic in the sense of Maldacena, Shenker and Stanford
 \cite{MalSheSta}. 
 Hence the SYK model, although very simple,
 offers a path to the main theoretical concepts of quantum gravity,
 such as Bekenstein-Hawking entropy and holography. 

SYK became a very active field, from the early papers to nowadays.
Quite naturally we devoted our common article with Nicolas Delporte 
to that subject and we entitled it  ``Holographic Tensors" \cite{tensortrack5}.

At large $N$ the Schwinger-Dyson equation for the 2-point function is closed. The conformal  symmetry 
can be broken and  the corresponding subject goes under the name of
near-$AdS_2$/near-$CFT_1$ correspondence. 
This  entails a relationship with Jackiw-Teitelboim two-dimensional  quantum gravity.

Witten has found a genuine field theory model (with no disorder), 
 in which the tensors plays a much more fundamental role \cite{witten2019syk}. 
In a nutshell, he discovered   that his model has the same melonic limit as the tensors models pioneered by Gur\u{a}u.
Klebanov and Tarnopolsky \cite{klebanov2017uncolored}, when combined with an earlier work of Carrozza and Tanasa \cite{carrozza2016n},
allows on a big simplification of the group symmetry of the main tensor, from $U(N)^{D(D-1)/2}$ to $O(N)^{D}$.

Unlike the initial SYK model, the field tensor models of  \cite{witten2019syk,klebanov2017uncolored,carrozza2016n}
fit in the framework of local quantum field theory with $D=1$.
Hence there is a possibility to extend them in $D>1$! But these models of SYK-type still are 
quantum mechanical and lost background-independence since they make use of a preferred time.
We stumbled for a while to that particular problem,
namely to restore background-independence to models of SYK-type.
Zero dimensional tensor models create naturally trees or unicycles as Gromov-Hausdorff
limits. If we can approximate the sub-dominant terms as matter fields living on
trees or unicycle (and it's a big ``if''), we shall get in this approximation an SYK-type model on a random tree or unicycle.
Thermal Euclidean, which plays such a natural role in SYK models, leads us to choose unicycles rather than trees.
Models of this type can be studied first by perturbative field theory techniques.

Then our main result together with Nicolas Delporte \cite{tensortrack6,DelRiv} is that, under reasonable assumptions, the SYK model for bosons averaged on long, infrared unicycles possesses a two-point function exhibiting  much the same behavior, but with a critical infrared exponent different from the one of ordinary SYK, sensitive to the spectral dimension of the underlying graph.

This can be seen as a simpler version of the well-known
2d CFT coupled to gravity (ie CFT on $\mathbb{R}^2$ but coupled to the Liouville field). The change in critical
exponent is a simpler analog of the Knizhnik-Polyakov-Zamolodchikov and David-Distler-Kawai relations, which tell how 
critical exponents change when coupled to 2d gravity.
The cycle in a unicycle can be identified to a lattice-regularized flat $U(1)$ thermal circle, and the 
trees decorating the unicycles are then the unidimensional analog of the bidimensional Liouville field 
bumps which are at the source of the modification of critical exponents.
In this way, field theory  on random unicycles can be seen as ``gravity in one dimension" or ``gravitational time".

We return to a more general problem with Ouerfelli and Tamaazousti; that of making
a big (maybe too ambitious?) jump, from quantum gravity to artificial intelligence \cite{tensortrack7}. More specifically the context of that paper 
is the following. Tensor PCA was introduced in the pioneer work of \cite{RicMon} and consists in recovering a signal spike $v_0^{\otimes k}$ that has been corrupted by a noise tensor $Z$: $T =Z + \beta v_0 ^{\otimes k}$ where $v_0$ is a unitary vector and $\beta$ the signal to noise ratio.

Matrix data analysis and principal component analysis (PCA) is mostly stated in the  ``quantum-mechanics" language of eigenvalues rather than in the ``quantum-field theoretic" language of  invariants and (Feynman) graphs. For  tensors the quantum-field language is the natural one. An important task in tensor data analysis is therefore to translate
the results of matrix data analysis and PCA into the  quantum-field theoretic language of invariants and graphs.

An original connection have been made in \cite{OueTamRivI} between tensorial data analysis and the tensor track.
This connection is based on the introduction of matrices that are built out of a graph and  cutting an edge.
Indeed, given a graph invariant, we call ``cutting an edge" the fact of not performing a sum over the index associated to this edge, which gives us a matrix.
The eigenvector associated to the largest eigenvalue of this matrix can then be proven to be correlated to the signal vector $v$ for a significant range of signal-to-noise ratio $\beta$.

Another article with Ouerfelli and Tamaazousti refers to an heuristic algorithm named SMPI \cite{OueTamRivII}. 
This algorithm seems to be 
considerably better that the state of the art; more progress is expected during the coming months and years.

\section{Recent work from our group} 

Rasvan Gur\u{a}u is busy by combining random tensors and conformal field theory, and he performed recently a beautiful set of lectures  
entitled ``From random tensors to tensor field theory" in a six-weeks program at Institut Henri Poincaré  in the winter of 2023.
From his recent output I have extracted three papers that I consider emblematic of his activity in Heidelberg.

With Dario Benedetti, Sabine Harribey and Davide Lettera, he consider the long-range bosonic $O(N)^3$ model  
where $N$ gets large enough \cite{BenGurHarLet}. The model displays four large-$N$ fixed points and the authors confirm  that the $F$-theorem holds in this case. This 
result is subtle,  as one of the couplings (the ``tetrahedral" coupling) is imaginary, and therefore the model is non-unitary at finite $N$. 

In the same vein, with his collaborators J\"urgen Berges and Thimo Preis, he studied further the quantum field theory with global $O(N)^3$ symmetry where $N$ is sufficiently large.
He find that both asymptotic freedom and boundedness from below can be realized in this theory  \cite{BerGurPre}. 
To uncover its scale dependence, the authors analyze the renormalization group flow for the tensor field which features two real quartic couplings, $g_1$ and $g_2$ (which correspond to the pillow and to the double-trace), and an imaginary tetrahedral coupling $ig$ (it is here that the link with \cite{BenGurHarLet} is used). It turns out that all the couplings exhibit asymptotic freedom,
leading to a just-renormalizable field theory with an ultraviolet limit and a strongly coupled infrared limit in four space-time Euclidean dimensions, just like QCD! 

On the other hand, from his ``constructive vein" and with his collaborators Dario Benedetti, Hannes Keppler and Davide Lettera, he has begun a study of Ecalle's trans-series, again on the $O(N)$ model, but this time at small $N$. They recover that both the partition function $Z(g,N)$ and $W(g,N)=\log Z(g,N)$ are Borel summable functions.
Then,  using our constructive field theory techniques such as the loop vertex expansion \cite{RivLVE}, they prove that the trans-series expansion of the Taylor coefficients of these expansions, $Z_n(g)$ and $W_n(g)$, are different. What I find especially strong in their article is the fact that they were able to extend the Borel transform to the angle $\frac{3}{2}\pi$ and to prove that, while $W(g,N)$ displays contributions from arbitrarily many multi-instantons, $W_n(g)$ exhibits contributions of only up to $n$-instanton sectors.

Recently Razvan and I made a review entitled ``Quantum Gravity and Random Tensors"  in the context of the Poincaré Seminar \cite{GurRiva}.

Joseph Ben Geloun has defended his  french ``Habilitation à diriger des recherches". Along the many gems of his HdR, I choose one which is particularly spectacular. With Sanjaye Ramgoolam they have been able to give a combinatorial interpretation of the Kronecker coefficients, a problem which existed since 85 years! Kronecker coefficients
are widely studied in mathematics from many points of view: symmetric polynomials, complexity theory, combinatorics...
The standard mathematical approach is to think of them as  the structure constants
of irreducible representations of symmetric groups and it's around  this approach that they've built their results \cite{GelRamI}. 

Joseph Ben Geloun is a most active researcher and we would like to stressed four recent contributions of him.
With Dina Andriantsiory and Mustapha Lebbah, he propose a method of clustering for 3-order tensors of different dimensions  via an affinity matrix. Based on a notion of similarity between the tensor slices and the spread of information of each slice, their model builds an matrix on which they apply advanced clustering methods. The combination of all clusters 
delivers the desired multiway clustering. Their method and their associated algorithm, which they baptized MCAM, achieves competitive results compared with other known algorithms, both on  synthetics and real datasets \cite{AndGelLebI,AndGelLebII,AndGelLebIII}.

With Reiko Toriumi, he wants to escape the branched polymer phase of tensor models. 
They explore two just-renormalizable quartic enhanced\footnote{Enhanced tensor field theories
have dominant graphs that do not correspond to  melonic graphs.} tensor field theories  \cite{GelTor},
and they obtain results which I find interesting. 
At all orders, both models have a constant wave function renormalization
and therefore no anomalous dimension. They also analyse their RG flow, which depend of two coupling constants.
They compute the perturbative $\beta$-functions of these coupling constants at one-loop. 
For the first model, the flow exhibits neither asymptotic
freedom nor the ordinary Landau ghost of $\phi^4_4$: one of the coupling  stays fixed, and the other
has a linear behavior in the time scale.
For the second model both couplings do not flow at all, according this one-loop approximation.

In the article with  Andreas Pithis and Johannes Th\"uringen \cite{GelPitThu}, 
their goal is to find a phase transition from discrete quantum-gravitational to continuum geometry starting with tensor invariants.
 In the so-called cyclic-melonic potential approximation of a tensorial field theory on the $r$-dimensional torus,
Andreas Pithis and Johannes Th\"uringen recently showed, using functional renormalization group techniques, that no such phase transition 
is possible. In \cite{GelPitThu} they show how to overcome this limitation by introducing local degrees of freedom on ${\mathbb R}^d$.
They find that the effective $r-1$ dimensions of the torus part dynamically vanish along the renormalization group flow 
while the $d$ local dimensions persist up to small momentum scales. Consequently, for $d>2$ they find a possibility to allow some phase transitions.

Recently Joseph Ben Geloun and Sanjaye Ramgoolam propose a theory of complexity pertaining to data analysis \cite{GelRamII}.
Their goal is to detect projectors in associative algebras, labelled by representation data. 
To illustrate this theory they propose three examples. One is based on a quantum algorithm 
based on quantum phase estimation, and they compare it to a classical  algorithm based on the AdS/CFT correspondence.
The other two, around the line of \cite{GelRamI}, are projectors labelled by a triple of Young diagrams,  all having $n$ boxes, 
or with $m,n$ and $m+n$ boxes, and having non-vanishing Kronecker coefficients  in the case of having $n$ boxes, 
or having non-vanishing Littlewood-Richardson coefficients in the case of having $m,n$ and $m+n$ boxes.

Adrian Tanasa is very active in Bordeaux. His book on combinatorial physics has been published \cite{Tan} 
and his scientific interests often revolve around the problem of  the double scaling of many models \cite{BonNadTan,BonNadTanII,KraMulTan}. 
With Hannes Keppler, Thomas Krajewski and Thomas Muller, he is also interested in so called negative dimension theorems, or $N$ to $-N$ 
dualities, relating the orthogonal and symplectic group vial the formal relation $SO(-N)\simeq Sp(N)$ \cite{KepKraMulTan}.

Sabine Harribey is currently a postdoc in Nordita. With her, Igor Klebanov and Zimo Sun
explore a new approach to boundaries and interfaces in the $O(N)$ model where they add certain localized cubic interactions. 
They use the technique of $1/\epsilon$ expansion and
they show that the one-loop beta functions of the cubic couplings are affected by the quartic bulk interactions. For the interfaces, they find real fixed points up to the critical value $N_{\rm crit}\approx 7$, while for $N> 4$ there are IR stable fixed points with purely imaginary values of the cubic couplings
\cite{HarKleSun}. Recently with Dario Benedetti, Razvan Gurau and Davide Lettera, she publishes an article on finite-size versus finite-temperature effets in the O(N) model
\cite{DarHarGurLet2}. In it they consider the classical model, which is conformally invariant at criticality, and they introduce one compact spatial direction. They show that the finite size dynamically induces an effective mass and they compute the one-point functions for bilinear primary operators with arbitrary spin and twist. Second, they study the quantum model, mapped to a Euclidean anisotropic field theory, local in Euclidean time and long-range in space, which the authors dub fractional Lifshitz field theory. They show that this model admits a fixed point at zero temperature,

Dario Benedetti and Valentin Bonzom have both defended their HDR. Dario Benedetti has defended his HdR in the Ecole Polytechnique \cite{BenI}
and Valentin Bonzom  has defended his HdR in the Paris Nord university \cite{Bon}. 
Dario Benedetti, Sylvain Carrozza, Reiko Toriumi and Guillaume Valette studied the double and triple-scaling limits of a complex multi-matrix model.
Their main result is, in a double-scaling limit, to characterize the Feynman graphs of arbitrary genus,
and in the triple-scaling limit,  to classify all the three-edge connected dominant graphs and to prove that their critical behavior falls in  the  universality  class  of Liouville  quantum  gravity \cite{BenCarTorVal}.

Sylvain Carrozza has been recruited permanently in Université de Bourgogne in Dijon.
We would like to stressed one recent contributions in his domain (apart from  \cite{BenCarTorVal}).
For $p = 3 $ and $p=5$ there exist a melonic large $N$ limit for $p$-irreducible tensors in the sense of Young tableaux. 
Sylvain Carrozza and Sabine Harribey overcome huge difficulties to solve the case $p=5$ \cite{CarHar}.
They demonstrate that random tensors transforming under rank-$5$ irreducible representations of $\mathrm{O}(N)$ can support melonic large $N$ expansions. Their construction is based on models with sextic ($5$-simplex) interaction, which generalize previously studied rank-$3$ models with quartic (tetrahedral) interaction. 
Their proof relies on recursive bounds derived from a detailed combinatorial analysis of the Feynman graphs. Their results provide further evidence that the melonic limit is a universal feature of irreducible tensor models in arbitrary rank. 

Luca Lionni has been recruited permanently in mathematics at Institut Camille Jordan in Lyon.
With Benoit Collins and Razvan Gur\u{a}u, he explore a generalization of the Harish-Chandra--Itzykson--Zuber integral to tensors and its expansion over trace-invariants of the two external tensors. This gives rise to natural generalizations of monotone double Hurwitz numbers.
They find an expression of these numbers in terms of monotone simple Hurwitz numbers
and they give an interpretation of their different combinatorial quantities in terms of enumeration of nodal surfaces \cite{ColGurLioI}.
Still with Benoit Collins and Razvan Gur\u{a}u, they analyze a two-parameter class of asymptotic scalings
when $N$, the size of the tensors, is large enough, uncovering several non-trivial asymptotic regimes. 
This study is relevant for analyzing the entanglement properties of multipartite quantum systems \cite{ColGurLioII}.

With Timothy Budd, his research is centered in random triangulations of manifolds, 
a goal which is central to the random geometry approach to quantum gravity. In case of the 3-sphere
 the pursuit is held back by serious challenges, including the wide open problem of enumerating triangulations.
First, they identify a restricted family of triangulations, of which the enumeration appears less daunting.
Then they prove that these are in bijection with a combinatorial family of triples of plane trees satisfying restrictions.
An important ingredient is a reconstruction of the triangulations from triples of trees that results in a subset of the so-called locally constructible triangulations.
Finally, several exponential enumerative bounds are deduced from the triples of trees and some simulation results are presented \cite{BudLio}.

Stephane Dartois with Camille Male and Ion Nechita studied
the tensor flattenings appear naturally in quantum information when one produces a density matrix by partially tracing the degrees of freedom of a pure quantum state. In their paper, they study the joint distribution of the flattenings of large random tensors under mild assumptions, in the sense of  free probability theory. 
They show the convergence toward an operator-valued circular system with amalgamation on permutation group algebras for which we describe the covariance structure. As an application they describe the law of large random density matrix of bosonic quantum states \cite{DarMalNec}. 

Nicolas Delporte has defended his PhD \cite{Del}. With Dario Benedetti he revisit the Amit-Roginsky (AR) model in the light of recent studies on SYK and tensor models.
It is a model of $N$ scalar fields transforming in an $N$-dimensional irreducible representation of $SO(3)$. The most relevant (in renormalization group sense) invariant interaction is cubic in the fields and mediated by a Wigner $3jm$ symbol. The latter can be viewed as a particular rank-3 tensor coupling, thus highlighting the similarity to the SYK model, in which the tensor coupling is however random and of even rank. As in the SYK and tensor models, in the large-$N$ limit the perturbative expansion is dominated by melonic diagrams. The lack of randomness, and the rapidly growing number of invariants that can be built with $n$ fields, makes the AR model somewhat closer to tensor. In the short range version of the model, they find, for $5.74<d<6$, a fixed-point defining a real CFT, while for smaller $d$ complex dimensions appear. They also introduce and study a long-range version of the model, for which the cubic interaction is marginal at large $N$, and they find a real and unitary CFT for any $d<6$, both for real and imaginary coupling constant, up to a critical coupling \cite{BenDel}.

Riccardo Martini and Reiko Toriumi give a procedure to construct 
trisection diagrams for closed pseudo-manifolds generated by colored tensor models without restrictions on the number of simplices in the triangulation, therefore generalizing previous works in the context of crystallizations and PL-manifolds. They further speculate on generalization of similar constructions for a class of pseudo-manifolds generated by simplicial colored tensor models \cite{MarTor}.

The authors of \cite{AbrPerTor} introduce a dually-weighted multi-matrix model that for a suitable choice of weights reproduce two-dimensional Causal Dynamical Triangulations (CDT) coupled to the Ising model. When Ising degrees of freedom are removed, this model corresponds to the 2d CDT-matrix model introduced by Dario Benedetti and Joe Henson
\cite{BenHen}. They present exact as well as approximate results for the Gaussian averages of characters of a Hermitian matrix $A$ and $A^2$ for a given representation and establish the present limitations that prevent them to solve the model analytically. This sets the stage for the formulation of more sophisticated matter models coupled to two-dimensional CDT as dually weighted multi-matrix models providing a complementary view to the standard simplicial formulation of CDT-matter models.

Vincent Lahoche together with Corinne de Lacroix and Harold Erbin \cite{LacErbLah} compute the gravitational action of a free massive Majorana fermion coupled to two-dimensional gravity on compact Riemann surfaces of arbitrary genus. The structure is similar to the case of the massive scalar. The small-mass expansion of the gravitational yields the Liouville action at zeroth order, and they can identify the Mabuchi action at first order. While the massive Majorana action is a conformal deformation of the massless Majorana CFT, they find an action different from the one given by the David-Distler-Kawai (DDK) ansatz.

Bio Wahabou Kpera, Vincent Lahoche and Dine Samary want to study the probability laws associated with random tensors or tensor field theories.
Their approach is to the quantize through a Langevin type equation. 
The method they propose use the self averaging property of the tensorial invariants in the large $N$ limit.
 In this regime, the dynamics is governed by the melonic sector.
Their work focuses on the cyclic (i.e. non-branching) melonic sector, 
and they study the way that the system returns to the equilibrium regime.
In particular, they provide a general formula for the transition temperature between these regimes. 
Numerical simulations are made to support the theoretical analysis
\cite{KpeLahSam}. Also they explored with Seke Yerima the Ward-Takahashi identities in tensorial group
field theories. Ward's identities result from a expansion around the identity,
and it is expected that a first-order expansion is indeed sufficient. 
They show that it doesn't occur for a complex tensor theory model with a kinetic term involving a Laplacian
\cite{KpeLahSamYer}. 

\section{Blitz Review of Stochastic Analysis}

Stochastic analysis has been revolutionized by the work of Martin Hairer on regularity structures \cite{Hai}, 
which provides a framework for studying a large class of stochastic partial differential equations arising from quantum field theory.
This framework covers the Kardar–Parisi–Zhang equation, the 
$ \Phi_{3}^{4}$ equation and the parabolic Anderson model, all of which require renormalization in order to have a well-defined notion of solution.
Among the fashionable follows-up we shall cite the dynamical approach of Barashkov and Gubinelli \cite{BarGub} and the variational approach 
of Gubinelli  and Hofmanová  \cite{GubHof}.

Regularity structures and associated models are a way of  solving a stochastic equation  
by choosing  objects replacing the polynomials of the Taylor expansion
with non-polynomial coefficients of the increase $h$ between $x$ and $x+h$. This generalized expansion 
is based on non-integer powers and even negatives ones.
In doing so, we perform intelligently a renormalization which 
depends only the form of the stochastic equation, not the constants involved  or  the stochastic noise. 

This point of view follows the footsteps of the fathers of differential calculus,
where $ \frac{ df}{dx}$ can make sense without having to make sense of $df$ and $dx$
separately, and the point of view of distributions,
where $\int \phi (x)\delta (x) dx$ makes sense for a test function $\phi$ without having to make sense of $\delta (x)$, the Dirac ``function".
But the later developments overcome the main obstacle of the latter theory, the multiplication of distributions!

Like Wilson's, Hairer's point of view is susceptible to many generalizations. Let's start with a trivial example:
$$A_\epsilon  = \{{[x_\epsilon (t), y_\epsilon (t)}]\,, t\in \mathbb{R}\},\qquad
x_\epsilon (t) = \epsilon  t + \frac{1}{\epsilon } ,\quad y_\epsilon (t) = \epsilon  \cos(t) .$$
At $t$ fixed and $\epsilon \to 0$, ${[x_\epsilon (t) , y_\epsilon (t)}] \to {[\infty, 0}]$,  
does not converge. If we perform a reparametrization $t \mapsto t/\epsilon  - 1/\epsilon^2$,
the solution reparametrised
$$
{[\hat x_\eps(t) = t \ , \hat y_\eps(t) = \eps \cos (\frac{t}{\eps} - \frac{1}{\eps^2})}] \to
 A_0 = \R \times \{0\}$$  indeed converges. In this analogy $(x_\eps,y_\eps)$ 
plays the role of the 'bare' solution $\phi_\eps$, while
 $(\hat x_\eps,\hat y_\eps)$ plays the role of the renormalized solution.
 
Let us come the heart of Hairer's formalism. 

\begin{definition}
A regularity structure is a triple ${\mathcal {T}}=(A,T,G)$ consisting of:
\begin{itemize}
\item a subset $A$ (index set) of $\mathbb {R}$  that is bounded from below and has no accumulation points;

\item the model space: a graded vector space $ T=\oplus _{\alpha \in A}T_{\alpha }$, where each $T_{\alpha }$ is a Banach space; 

\item and the structure group: a group $G$ of continuous linear operators $\Gamma \colon T\to T$ such that, for each $ \alpha \in A$
and each $ \tau \in T_{\alpha }$, we have 
\bea
 (\Gamma -1)\tau \in \oplus _{\beta <\alpha }T_{\beta } . 
 \label{eqA}
\eea
  Moreover $ \Gamma 1 = 1 $ for each $ \Gamma \in G $.
\end{itemize}
\end{definition}
Let us use the notation $X^k$ for $X_1^{k_1}\cdots X_d^{k_d}$ for any multi-index $k$.
The polynomial canonical model we're trying to generalize is then given by
\begin{itemize}
\item $A = \N$,

\item $T_k $ the linear space generated by monomials of degree $k$,

\item a group of structure $G$ (in this case $\R^d$ endowed with addition) 
which acts on $T$ via $\Gamma_h X^k = (X-h)^k$, $h \in \R^d$.

\end{itemize}
However, it's not the complete algebraic structure that describes Taylor developments. 
An essential element is that a development around a certain point $x_0$ can be 
developed around any other point $x_1$ by the formula
\begin{equation}
(x-x_0)^m = \sum_{k+\ell = m} \binom{m}{k} (x_1 - x_0)^k\cdot (x-x_1)^\ell\;. \nonumber
\end{equation}
Such a redevelopment application
$\Gamma_{st}$ has the property that
\begin{eqnarray}
\{ \Gamma_{st} \tau - \tau\} \ \in \bigoplus_{\beta < \alpha} T_\beta = : T_{<\alpha}.  \nonumber
\end{eqnarray}
In other words, when redeveloping a homogeneous monomial around a different point, 
the leading-order coefficient remains the same, but lower-order monomials may appear.
This is the meaning of the condition \eqref{eqA}.

 With the algebraic skeleton so defined, we move on to the associated analytical flesh. 
A further key notion is that of a \emph{model}, which is a 
way of associating to any $\tau \in T$ and $x_{0}\in \mathbb {R} ^{d}$ a ``Taylor polynomial" based at 
$x_{0}$ and represented by $\tau$, subject to some consistency requirements.
Concretely it consisting of a \emph{family of applications} 
 \bea
\Pi \colon \R^d &\to \CL\bigl(T, \CS'(\R^d)\bigr);&\qquad \Gamma\colon \R^d \times \R^d \to G\   \nonumber
\eea
where $\CS'(\R^d)$ is the space of distributions (not necessarily temperate) 
on $\R^d$. In the polynomial natural canonical model this family of applications is
\bea
\bigl(\Pi_x X^k\bigr)(y) = (y-x)^k,\quad \Gamma_{xy} = \Gamma_{y-x}, \quad \Gamma_{xy}\Gamma_{yz} = \Gamma_{xz}, \quad  \Pi_x \Gamma_{xy} = \Pi_y. \nonumber
\eea

A first difficulty arises. We want to allow $\tau$ elements in $T_\alpha$ to represent distributions and not just functions that
cancel each other out to order $\alpha$ around a point. But then we cannot evaluate them point by point.

A second difficulty lies in the notion to ``cancel to the order of $\alpha$" when
$\alpha$ is negative. It can only say tend to infinity slowly enough as a function of $\alpha$.
We therefore need a
extended notion of ``cancelling to the order of $\alpha$". We will achieve this by controlling
the size of our distributions in a small region around the given point $x_0$.
Consider a test function $\phi$ and define
\bea
\phi_x^\lambda(y) := \lambda^{-d} \phi\bigl(\lambda^{-1}(y-x)\bigr)\ . \nonumber
\eea

\def\CC{{\mathcal C}}
Let $r $ be a natural number and $\CB_r$ the set of
test functions $\phi \colon \R^d \to \R$ with support in the ball $B(0,1)$
such as $\|\phi\|_{\CC^r} \le 1$.
Combining intelligently everything, we can define at last the notion of models:
\begin{definition}
Given a regularity structure ${\mathfrak T}$ and an integer $d \ge 1$, a \emph{model} for ${\mathfrak T}$ over $\R^d$ consists
applications
\bea
\Pi \colon \R^d &\to& \CL\bigl(T, \CS'(\R^d)\bigr)\qquad \Gamma\colon \R^d \times \R^d \to \; G\ \nonumber\\
x\quad &\mapsto& \qquad \Pi_x\quad \quad \quad \quad\quad\quad (x,y) \ \ \mapsto\; \Gamma_{xy} \nonumber
\eea
so that $\Gamma_{xy}\Gamma_{yz} = \Gamma_{xz}$ and $\Pi_x \Gamma_{xy} = \Pi_y$ (Chen relations). Moreover, given $r > |\inf A|$,
for any compact set $\K \subset \R^d$ and constant $\gamma > 0$ there exists a constant $C$ such that the bounds
\begin{equation}
\bigl|\bigl(\Pi_x \tau\bigr)(\phi_x^\lambda)\bigr| \le C \lambda^{|\tau|} \|\tau\|_\alpha\;,\qquad
\|\Gamma_{xy}\tau\|_\beta \le C |x-y|^{\alpha-\beta} \|\tau\|_\alpha\;, \nonumber
\end{equation}
are satisfied uniformly over $\phi \in \CB_r$, $(x,y) \in \K$, $\lambda \in (0,1]$,
$\tau \in T_\alpha$, with $\alpha \le \gamma$ and $\beta < \alpha$.
\end{definition}

Let us make a few remarks.
\begin{itemize}

\item
Chen's relations $\Gamma_{xy}\Gamma_{yz} = \Gamma_{xz}$ and $\Pi_x \Gamma_{xy} = \Pi_y$ are natural. 

\medskip\item
The first bound indicates precisely what we mean when we say that $\tau \in T_\alpha$ represents a term 
of the order of $\alpha$.

\medskip\item
The second bound is also very natural: it indicates that 
when developing a monomer of order $\alpha$ around a new point 
at a distance $h$ from the old one,
the coefficient in front of monomials of order $\beta$ is at most of order $h^{\alpha - \beta}$.

\medskip\item
In many interesting cases, it's natural to scale the different directions of $\R^d$ in a different way. 
For example to construct solutions for stochastic parabolic PDEs, where
the time direction ``counts double",  we define $\phi_x^\lambda$ so that the $i$th
direction is dilated by $\lambda^{s_i}$, where $s=\{ s_i , i= 1,\cdots , d \}$.
In this case, the corresponding scaled distance is 
$|x|_s = \sum_i |x_i|^{1/s_i}$.

\medskip\item
For a given structure and regularity pattern, the distribution represented by $f$ should be $\CR f(x) := [\Pi_x f](x)$.
However, this definition is not correct since $\Pi_x f(x)$ is in general a distribution, and therefore may not be evaluated in $x$!
In addition the models must be subject to a ``reconstruction theorem" which says that in a unique way
the distribution is  reconstructible by its generalized Taylor expansion.
And this theorem cannot be obtained in a \emph{linear} way.

\end{itemize}

Hairer's theory solves this difficulty by defining another space that Hairer calls ${D^\gamma}$,
endowed with a topology metrizable but not normable which
reflects the non-linearity of the problem.
\begin{definition}
Consider a regularity structure $\TT$ equipped with a model $(\Pi,\Gamma)$ defined on $\R^d$.
The space $\D^\gamma = \D^\gamma(\TT,\Gamma)$ is given by the set of functions
$f\colon \R^d \to \bigoplus_{\alpha < \gamma} T_\alpha$
such that, for each compact set $\K$ and each $\alpha < \gamma$, there exists a constant $C$ with
\bea
\| f(x) - \Gamma_{xy} f(y) \|_{\alpha} \le C |x-y|^{\gamma-\alpha}
\eea
uniformly over $x,y \in \K$.
\end{definition}
Let us comment on this definition.
$\MM \ltimes \CD^\gamma := \{
(\Pi,\Gamma,F)\}$ is a metrizable space but not a normed space.
The distance between $(\Pi, \Gamma, f)$ and $(\bar \Pi, \bar \Gamma, \bar f)$ is given by $\inf \rho$ such that
\bea
 \| f (x) - \bar f(x) - \Gamma_{xy} f(y) + \bar \Gamma_{xy} \bar f(y)\|_{\alpha} &\le& \rho |x-y| ^{\gamma-\alpha}, \nonumber\\
 \vert \bigl(\Pi_x \tau - \bar \Pi_x \tau\bigr) (\phi_x^\lambda) \vert &\le& \rho \lambda^{\alpha} \|\tau\| , \nonumber \\
\|\Gamma_{xy}\tau - \bar \Gamma_{xy}\tau\|_{\beta} &\le& \rho |x-y|^{\alpha-\beta} \|\tau\|\ , \nonumber
\eea
uniformly for $x,y$ in a compact set.
So we can formulate the reconstruction theorem:
\begin{theorem}
Let ${\mathfrak T}$ be a regularity structure as above and let $(\Pi,\Gamma)$ be a model for ${\mathfrak T}$ over $\R^d$.
There is a unique linear map $\CR\colon \D^\gamma \to \CS'(\R^d)$ such that
\begin{equation}
\bigl|\bigl(\CR f - \Pi_x f(x)\bigr)(\phi_x^\lambda)\bigr| \lesssim \lambda^\gamma\ , \label{rebuild}
\end{equation}
uniformly over $\phi \in \CB_r$ and $\lambda$ and locally uniformly over $x$.
\end{theorem}

Uniqueness is much easier than existence. The existence of such a function
appealed in a crucial way to the existence of a wavelet basis
consisting of $C_r$ functions with compact support, which was demonstrated in 1988
by Ingrid Daubechies.

We take as a dynamic point of view the example of $\Phi^4$ which stochastic equation is
\begin{equation}
\partial_t \Phi = \Delta \Phi - \Phi^3 + \xi\;, \label{phi4}
\end{equation}

\begin{itemize}
\item
We place ourselves in dimension 3 on a torus $T^3$ in direct space,

\item
 $\xi$ is a stochastic variable, for example Gaussian with covariance
\begin{equation}
{\mathbf E} \xi(t,x)\xi(s,y) = \delta(t-s)\delta(y-x)\;.
\end{equation}

\item
 $\rho_\eps$ is a
``mollifier" smooth with compact support which tends when $\eps \to 0$ to $\delta(t-s)\delta(y-x)$,

\item the \emph{smoothed} stochastic equation $u_\eps$ for $\xi_\eps = \rho_\eps * \xi$,
\begin{equation}
\partial_t u_\eps = \Delta u_\eps + C_\eps u_\eps - u_\eps^3 + \xi_\eps \; ,
\end{equation}
should tend to the sought solution when $\eps \to 0$.
\end{itemize}

\begin{figure}[h!]
\includegraphics[scale=0.76,angle=270]{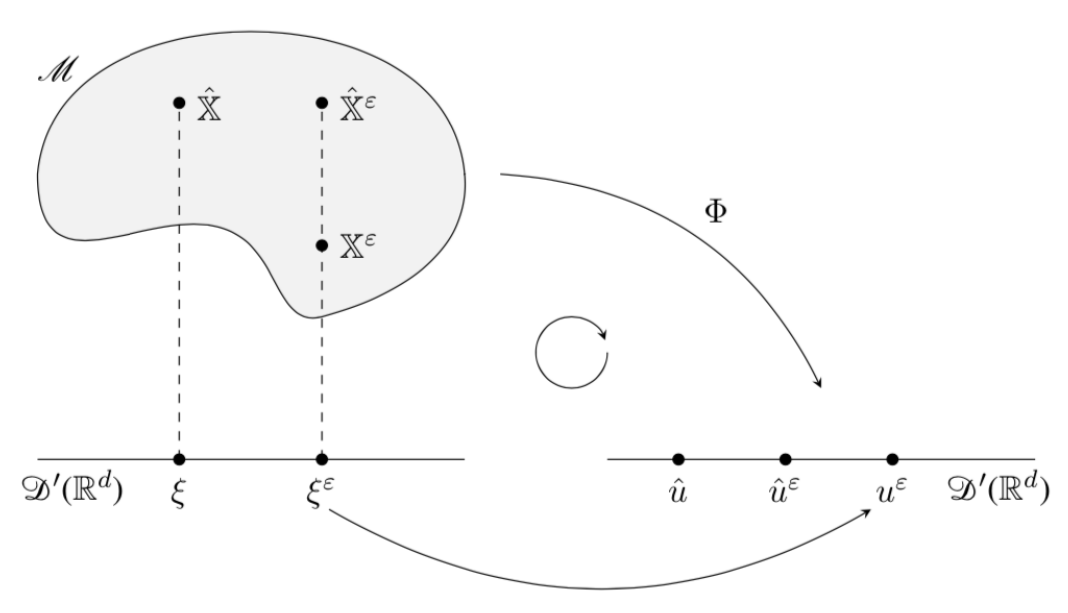} 
\end{figure}
In this figure, we show the factorization of the application $\xi^\gep\mapsto u^\gep$ into
$\xi^\gep\mapsto{\mathbb X}^\gep\mapsto \Phi({\mathbb X}^\gep)=u^\gep$. We can also see that in the space of $\MM$ models we have several antecedents 
of $\xi^\gep\in{\mathcal S}'(\R^d)$, for example the canonical model ${\mathbb X}^\gep$ and the renormalized model $\hat 
{\mathbb X}^\gep$; it's the latter that converges to a $\hat{\mathbb X}$ model, thus offering a bearing on $\xi$. Note that $\hat u^\gep=\Phi(\hat{\mathbb X}^\gep)$ and $\hat u=\Phi(\hat{\mathbb X})$.

The BPHZ solution goes through the following steps

\medskip
\begin{itemize}
\item {\it Algebraic step}: Construction of the space of models $(\MM,{\rm d})$ and
 renormalization of the canonical model $\MM\ni{\mathbb X}^\gep\mapsto\hat{\mathbb X}^\gep\in\MM$ \cite{BruHaiZam}.
\medskip\item {\it Analytical step}: Continuity of application $\MM\to\CS'(\R^d)$ \cite{Hai}.
\medskip\item {\it Probabilistic step}: Convergence in the sense of probabilities
from the renormalized model $\hat{\mathbb X}^\gep$ to $\hat{\mathbb X} \in (\MM,{\rm d})$ \cite{ChaHai}.
\medskip\item {\it Second algebraic step}: Identification of the final application $\Phi(\hat{\mathbb X}^\gep)$
with the classical solution with local counterterms at the BPHZ \cite{BruChaCheHai}.
\end{itemize}

\section{The work of Léonard Ferdinand}

In this section we summarize briefly the articles of Léonard Ferdinand, a PhD of mine.

Léonard Ferdinand, Razvan Gur\u{a}u, Carlos Perez-Sanchez and Fabien Vignes-Tourneret 
consider a quartic $O(N)$-vector model \cite{FerGurPerVig}. Using the loop
vertex expansion, they prove the Borel summability in $1/N$ for the cumulants (including the free energy, which one considers the cumulant of zero order). The Borel summability holds uniformly in the coupling constant, as long as the latter belongs to a cardioid domain of the complex plane. Among their toolbox, let's remark that they use ciliated trees in a relatively new sense.

Then we summarize briefly the articles of Ismail Bailleul, Nguyen Viet Dang, Léonard Ferdinand and Tat Dat Tô 
\cite{BaiDanFerToI,BaiDanFerToII}. These authors deal with the $\Phi^4_3$ measure, but 
what is more original, on a arbitrary 3-dimensional compact Riemannian manifold without boundary.
They prove the nontriviality and the local covariance under Riemannian isometries of the corresponding measure.
This answers a longstanding open problem of constructive quantum field theory on curved backgrounds in dimension 3. 
To control analytically several Feynman diagrams appearing in the construction of a number of random fields, 
they introduce a novel approach of renormalization using microlocal and harmonic analysis. 
This allows them to obtain a renormalized equation which involves some universal constants independent of the manifold. 

In a companion paper \cite{BaiDanFerToII}, they develop in a self-contained way all the tools from paradifferential 
and microlocal analysis that they use in \cite{BaiDanFerToI}, setting a number of analytic and probabilistic objects.
In \cite{BaiDanFerLecLin} the authors argue that the spectrally cut-off Gaussian free field $\Phi_\Lambda$ 
on a compact riemannian manifold or on $\mathbb{R}^n$ cannot satisfy the spatial Markov property.

Finally we want to summarize the article of Ajay Chandra and Léonard Ferdinand
\cite{ChaFer}. These authors present two different arguments using stochastic analysis to construct 
super-renormalizable tensor field theories, namely the $T^4_3$ and $T^4_4$ models. 
The first approach is the construction of a Langevin dynamic \cite{Hai,GubHof} combined with a 
PDE energy estimate while the second is an application of the variational approach of Barashkov and Gubinelli \cite{BarGub}. 
By leveraging the melonic structure of divergences, regularising properties of non-local products, and controlling 
certain random operators, they demonstrate that for tensor field theories these arguments 
can be significantly simplified in comparison to what is required for $\Phi^4_3$ model. 

In their most recent article \cite{ChaFer1} Ajay Chandra and Léonard Ferdinand show that the flow approach of Duch [Duc21] can be adapted to prove local well-posedness for the generalised KPZ equation. 
The key step is to extend the flow approach so that it can accommodate semi-linear equations involving smooth functions of the solution instead of only polynomials - 
this is accomplished by introducing coordinates for the flow built out of the elementary differentials associated to the equation.

\section{Conclusion}

This review is a modest step in the direction of bring closer the different people working on random tensors and stochastic analysis
and it suggests research in many directions, among which:

 \medskip
\begin{itemize}

\item our next goal is $T^4_5$ which is just renormalizable and asymptotically free. We are reasonably confident that it can be solvable soon among the strategy
and the tactics defined by \cite{ChaFer1,RivVig},

\item for the future one should further develop the model of Razvan Gurau and collaborators \cite{BerGurPre} in the direction of constructive field theory.
This promising model lies in the class of tensor field theory with imaginary tetrahedral coupling, 
and, what is more important for the physics,  it is asymptotically free and
in dimension four,

\item in the more distant future, the common goal of all  the people working on field theory with 
stochastic analysis (including Martin Hairer himself \cite{ChaCheHaiShe}) is to tackle \emph{gauge theories} in the spirit of Parisi-Wu \cite{ParWu}.
\end{itemize}

\section{Appendix: Besov spaces}
\label{app:besov}
\newcommand{\T}{{\mathbb T}}

In mathematics, the Besov space 
$B_{p,q}^{s}(\mathbf {R} )$ is a complete quasinormed space which is a Banach space when $1 \leq p, q \leq \infty$. These spaces serve to generalize more elementary function spaces such as Sobolev spaces and are effective at measuring regularity properties of functions.

Let $ \Delta _{h}f(x)=f(x-h)-f(x)$ and define the modulus of continuity by
\bea \omega _{p}^{2}(f,t)=\sup_{|h|\leq t}\left\|\Delta _{h}^{2}f\right\|_{p}.
\eea
Let $n$ be a non-negative integer and define:
$s = n + \alpha$ with $0 < \alpha \leq 1$. The Besov space $ B_{p,q}^{s}(\mathbf {R} )$ contains all functions $f$ such that
\bea f\in W^{n,p}(\mathbf {R} ),\qquad \int _{0}^{\infty }\left|{\frac {\omega _{p}^{2}\left(f^{(n)},t\right)}{t^{\alpha }}}\right|^{q}{\frac {dt}{t}}<\infty .
\eea
The Besov space $B_{p,q}^{s}(\mathbf {R} )$ is equipped with the norm
\bea \left\|f\right\|_{B_{p,q}^{s}(\mathbf {R} )}=\left(\|f\|_{W^{n,p}(\mathbf {R} )}^{q}+\int _{0}^{\infty }\left|{\frac {\omega _{p}^{2}\left(f^{(n)},t\right)}{t^{\alpha }}}\right|^{q}{\frac {dt}{t}}\right)^{\frac {1}{q}}.
\eea
The Besov spaces $B_{2,2}^{s}(\mathbf {R} )$ coincide with the more classical Sobolev spaces 
$ H^{s}(\mathbf {R} )$.

Next, on a variety such as $B_{p,q}^{s}(\T^d)$, the authors of \cite{BaiDanFerToI} want to define a number of operators on functions spaces by using local charts. For that task they import some known regularity properties of the corresponding objects from the flat to the curved setting. They denote as usual by $B^\gamma_{p,q}(\T^d)$ the Besov spaces over $\T^d$ and by $C^\gamma(\T^d)$ the Besov-H\"older space $B^{\gamma}_{\infty,\infty}(\T^d)$, with associated norm denoted by $\Vert\cdot\Vert_{C^\gamma}$.
And so they define and use of the Besov spaces  in their analysis.

\end{document}